\documentclass[conference]{IEEEtran}

\usepackage{amsmath}
\usepackage{amsfonts}
\usepackage{amssymb}
\usepackage{epsf}
\usepackage{cite}
\usepackage{balance}
\usepackage{fancyhdr}
\usepackage{graphicx}
\usepackage{subfigure}
\usepackage[blocks]{authblk}
\usepackage[stable]{footmisc}
\usepackage{float}
\usepackage{fancyhdr}
\usepackage{gensymb}
\usepackage[bookmarks=false]{hyperref}
\usepackage{epstopdf}

\makeatletter
\def\ps@IEEEtitlepagestyle{%
  \def\@oddfoot{\mycopyrightnotice}%
  \def\@evenfoot{}%
}
\def\mycopyrightnotice{%
  {\footnotesize \copyright 2015 IEEE. Personal use of this material is permitted. Permission from IEEE must be obtained for all other uses, in any current or future media\hfill}
  \gdef\mycopyrightnotice{}
}

\newcounter{subeq}

\IEEEoverridecommandlockouts
\date{}
\begin{document}
\title{Multipath Reflections Analysis on Indoor Visible Light Positioning System}

\author {Wenjun Gu}
\author{Mohammadreza A. Kashani}
\author {Mohsen Kavehrad}

\affil{Department of Electrical Engineering\authorcr
The Pennsylvania State University, University Park, PA 16802\authorcr
Email: \{wzg112,mza159, mkavehrad\}@psu.edu\authorcr}
\maketitle
\begin{abstract}
Visible light communication (VLC) has become a promising research topic in recent years, and finds its wide applications in indoor environments. Particularly, for location based services (LBS), visible light also provides a practical solution for indoor positioning. Multipath-induced dispersion is one of the major concerns for complex indoor environments. It affects not only the communication performance but also the positioning accuracy. In this paper, we investigate the impact of multipath reflections on the positioning accuracy of indoor VLC positioning systems. Combined Deterministic and Modified Monte Carlo (CDMMC) approach is applied to estimate the channel impulse response considering multipath reflections. Since the received signal strength (RSS) information is used for the positioning algorithm, the power distribution from one transmitter in a typical room configuration is first calculated. Then, the positioning accuracy in terms of root mean square error is obtained and analyzed.
 \end{abstract}
\begin{IEEEkeywords}
Visible light communication, indoor positioning, multipath reflections, impulse response, received signal strength.
\end{IEEEkeywords}

\section{Introduction}\label{INTRODUCTION}
\IEEEPARstart{L}{ight-emitting-diodes} (LED) technology has been developing rapidly in recent years. It provides efficient and economical illumination as well as long service life time. Meanwhile, LED technology also finds its wide application in visible light communication (VLC) area as LEDs can be modulated at relatively high speeds \cite{1.1,1.2,1.3,peng,1.4,1}. For indoor environment, Global Positioning System (GPS) signal, serving for outdoor positioning, suffers from large attenuation when penetrates solid walls. VLC provides a practical solution to realize indoor positioning so that location based services (LBS) are available. These systems further offer significant technical and operational advantages. First, positioning systems based on VLC can be applied in many Radio Frequency (RF) sensitive environments as it does not add any electromagnetic interference. Second, employing the same infrastructure for both illumination and positioning considerably reduces development costs for industry.

For current indoor light positioning systems, LED bulbs act as the transmitters, and photo-diode (PD) collocated with a user is the receiver. Several positioning approaches have been proposed in the literature to calculate the receiver coordinates. In one approach, angulation algorithm is used to calculate the receiver position based on angle-of-arrival (AOA) information and rotation matrix \cite{2}. Scene analysis is another approach to obtain the receiver position. Features of each location are collected as the fingerprints in the offline stage. In the online stage, the features of current location are measured and by matching those with the offline fingerprints, the location of the receiver is estimated \cite{3}. In addition, Zigbee technology can be combined with VLC to realize long distance positioning \cite{4}. In this paper, we employ a commonly used algorithm where the received signal strength (RSS) information is first detected by PD, and then the distance between the transmitter and receiver is calculated. The lateration algorithm is finally applied to estimate the receiver coordinates \cite{5,6}. Gaussian mixture sigma point particle filter can be further employed to achieve high positioning accuracy and prevention of large deviations \cite{7}.

In the literature, line-of-sight (LOS) channels have been considered without taking into account the multipath reflections. However, the transmitted signal introduces multipath reflections as it bounces off the walls, ceiling and floor where the transmitter is a wide-beam LED source, and the receiver having a finite field-of-view (FOV) captures reflected photons from room surfaces. In this paper, we investigate the effect of multipath-induced distortion on the positioning accuracy of indoor light positioning systems. There are several methods to approximate the impulse response of an indoor optical wireless channel. In \cite{8}, Barry et al. proposed a deterministic algorithm that partitions a room into many elementary reflectors and sums up the impulse response contributions from different orders of bounces. As this method is recursive, it takes a significant amount of computation time.

Monte Carlo ray tracing approach is an alternative way to calculate the impulse response where rays with identical optical power are traced from the source \cite{9}. The direction of rays are generated by a probability density function (PDF) modeled by a Lambertian pattern. When the rays hit reflecting surfaces, new rays with reduced power are generated from the impact point with the same PDF. This method suffers from the fact that it needs a very large number of rays while only a small portion of rays will finally reach the PD. To alleviate this issue, modified Monte Carlo (MMC) approach is proposed which exploits each ray several times instead of only once \cite{10}. Although this method is fast, it introduces some variance due to the random direction of the rays. In \cite{11}, Alqudah and Kavehrad proposed a new approach to characterize diffuse links in a multiple-input multiple-output (MIMO) system. In this paper, we use the methodology developed recently in \cite{12} referred as combined deterministic and modified Monte Carlo (CDMMC), taking advantages of both methods to simulate the impulse responses of indoor optical wireless channels.

The rest of the paper is organized as follows. In Section II, the system model and CDMMC approach are briefly discussed. In Section III, positioning algorithm is described. In Section IV, we present computed numerical results of channel impulse response, power distribution and positioning accuracy. Finally, Section V concludes the paper.

\section{Multipath Analysis Approach}
\subsection{System Model}
\begin{figure}
\centering
\includegraphics[width = 8cm, height = 7.5cm]{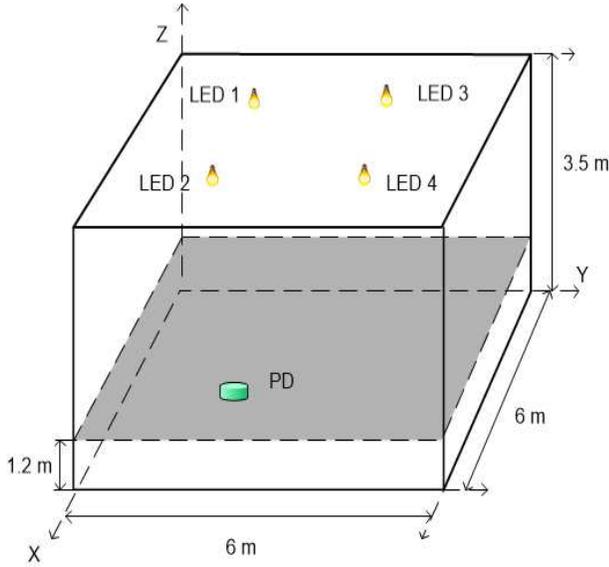}
\caption{System Configuration.}
\end{figure}

Fig. 1 shows the model of a typical 6 m $\times$ 6 m $\times$ 3.5 m indoor optical positioning system under consideration. We assume four LED bulbs are installed on the ceiling of a room, each of them acts as a transmitter. Each bulb has an identification (ID) code assigned to it which denotes its coordinates. These LED bulbs are modulated by the drivers with on-off-keying (OOK) format. As the bulbs are facing downwards, the elevation angle is -90$^{\circ}$ and azimuth angle is 0$^{\circ}$. Considering the installation, the height of the bulbs is 3.3 m. The Lambertian mode is order 1 and the wavelength is 420 nm. A receiver is located at the height of 1.2 m, with an area of 1$\times 10^{-4} \text{m}^{2}$  and field of view (FOV) of 70$^{\circ}$. As it is facing upwards, the elevation angle is +90$^{\circ}$ and the azimuth angle is 0$^{\circ}$. The corresponding reflection coefficients of the walls, ceiling and floor along with other parameters are summarized in Table 1.

\begin{table}
\begin{center}
\begin{quote}
\caption{SYSTEM PARAMETERS}
\label{table2}
\end{quote}
\begin{tabular}{|c|c|}
  \hline
  \textbf{Room dimensions} &\textbf{Reflection coefficients} \\ \hline
  length: 6 m &    ${{\rho }_{wall}}$: 0.66\\
  width: 6 m  & ${{\rho }_{Ceiling}}$: 0.35\\
  height: 3.5 m & ${{\rho }_{Floor}}$: 0.60\\ \hline
 \textbf{Transmitters (Sources)} &\textbf{Receiver} \\ \hline
  Wavelength: 420 nm &Area $\left(A\right)$: 1$\times 10^{-4} \text{m}^{2}$\\
  Height $\left(H\right)$: 3.3 m &Height $\left(h\right)$: 1.2 m\\
 Lambertian mode $\left(m\right)$: 1 &Elevation: +90$^{\circ}$\\
 Elevation: -90$^{\circ}$ &Azimuth: 0$^{\circ}$\\
 Azimuth: 0$^{\circ}$ &FOV $\left({{\Psi }_{c}}\right)$: 70$^{\circ}$\\
 Coordinates: (2,2) (2,4) (4,2) (4,4) & \\
 Power for "1"/ "0": 5 W/3 W & \\ \hline
\end{tabular}
\end{center}
\end{table}

\subsection{Impulse Response Analysis}
CDMMC method combines the deterministic and the MMC methods to take advantage of both. The contribution of first reflections to the total impulse response is calculated by the deterministic method, and then the contributions of second and higher-order reflections to the total impulse response are calculated by the MMC method.

In this method, room surfaces are first divided into small square elements, each of which has a size of  1$\times 10^{-4} \text{m}^{2}$. These elements and the PD are considered as the receivers. The received power is then obtained as
 \begin{equation}\label{eq1}
 {{P}_{received}}^{\left( 0 \right)}=H\left( 0 \right){{P}_{source}}^{\left( 0 \right)},
 \end{equation}
where $H\left( 0 \right)$ is the channel DC gain \cite{13}, and ${{P}_{source}}^{\left( 0 \right)}$ is the power from the LED bulbs provided in Table 1. With this step, the received power of LOS link is calculated.

Second, the small elements are considered as point sources which their source power can be calculated as
\begin{equation}\label{eq2}
{{P}_{source}}^{\left( 1 \right)}={{P}_{received}}^{\left( 0 \right)}{{\rho }_{surface}}.
\end{equation}
In Eq. (\ref{eq2}), ${{\rho }_{surface}}$ is the reflection coefficient of walls, ceiling or floor provided in Table 1.

Third, these small elements together with PD act as receivers again, and their received power is
\begin{equation}\label{eq3}
{{P}_{received}}^{\left( 1 \right)}=H\left( 0 \right){{P}_{source}}^{\left( 1 \right)}.
\end{equation}
With this step, the received power of first reflections is calculated.

To calculate the contribution of second reflections, 10 random rays from each of the elements are generated. The power ${{P}_{source}}^{\left(2\right)}$ of each ray is then expressed as
\begin{equation}\label{eq4}
{{P}_{source}}^{\left( 2 \right)}={{{P}_{received}}^{\left( 1 \right)}{{\rho }_{surface}}}/{10},
\end{equation}
\begin{figure}
\centering
\includegraphics[width = 7cm, height = 5cm]{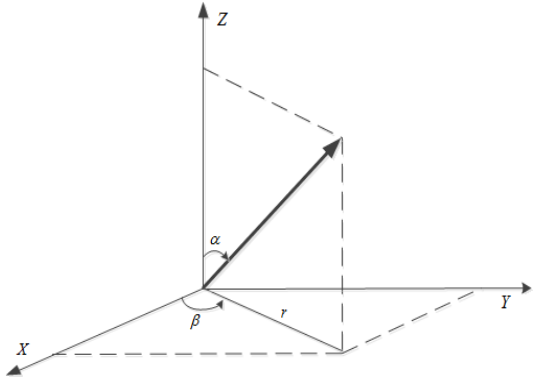}
\caption{Coordinate system for a random ray.}
\end{figure}
where the PDF of the rays' directions follows
\begin{equation}\label{eq5}
{f\left( \alpha, \beta\right)=\frac{m+1}{2\pi }{{\cos }^{m}}\left( \alpha  \right)}.
\end{equation}
In Eq. (\ref{eq5}), $m$ is the Lambertion order, $\alpha$ is the angle between the random ray vector shown in Fig. 2 and $z$-axis, and $\beta$ is the angle between $x$-axis and the vector's projection on the $X-Y$ plane. Note that Eq. (\ref{eq5}) is independent of $\beta$.

The $X-Y$ plane of the unit vector is the surface plane, and the origin point is the point source of each ray. These rays propagate through the room, and the contribution of the secondary reflections is obtained. When the rays hit the room surfaces, the impact points are considered as the new sources. New rays of each source are generated with reduced power where the PDF of their directions follows Eq. (\ref{eq5}). In this recursive approach, the contribution from subsequent reflections is calculated. A flow diagram of CDMMC algorithm is shown in Fig. 3. By summing up the contributions from different orders of reflections, the total impulse response of the channel is computed.
\begin{figure}
\centering
\includegraphics[width = 5cm, height = 7cm]{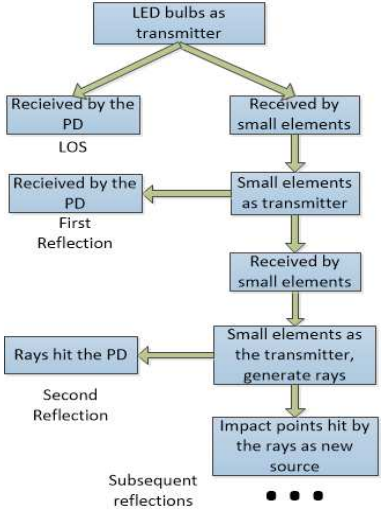}
\caption{Flow diagram of CDMMC approach}
\end{figure}
\section{POSITIONING ALGORITHM}
 After calculating impulse response with CDMMC method, the received signal strength from transmitter $i$ is obtained as $P_{r}^{\left( i \right)}$,  $i=1,2,3$ and 4. The distance ${{d}_{i}}$ between $i^\text{th}$ transmitter and receiver is then estimated according to the following equation:
\begin{equation}\label{eq6}
P_{r}^{\left( i \right)}=\frac{m+1}{2\pi {{d}_{i}}^{2}}A{{\cos }^{m}}\left( \phi  \right){{T}_{s}}\left( \psi  \right)g\left( \psi  \right)\cos \left( \psi  \right){{P}_{t}}.
\end{equation}
In particular, Eq. (\ref{eq6}) shows power attenuation from transmitter to receiver with respect to their distance where	${{P}_{t}}$ is the transmitted power. In (\ref{eq6}), Parameter $\phi$ is irradiance angle of transmitter, which numerically equals to the incident angle $\psi$ of the receiver and is obtained using $\cos \left( \psi  \right)=\cos \left( \phi  \right)=\left( H-h \right)/{{d}_{i}}$, where $H-h$ is the vertical distance between transmitter and receiver calculated based on the values listed in Table 1. ${{T}_{s}}\left( \psi  \right)=1$  is the transmittance function of the optical filter in the proposed system, and $g(\psi )$ is the gain for a compound parabolic concentrator (CPC) expressed as
\begin{equation}\label{eq7}
g\left( \psi  \right)=\begin{cases}
\frac{{{n}^{2}}}{{{\sin }^{2}}\left( {{\Psi }_{c}} \right)}, & 0\le \psi \le {{\Psi }_{c}}  \\
0, & \psi >{{\Psi }_{c}}  \\
\end{cases},
\end{equation}
where $n$ denotes the refractive index assumed as $n=1.5$, and ${{\Psi }_{c}}$ is the FOV provided in Table 1. Therefore, $d_{i}$ can be calculated as
\begin{equation}\label{eq8}
{{d}_{i}}=\sqrt{\frac{A{{T}_{s}}\left( \psi  \right)g\left( \psi  \right){{P}_{t}}{{\left( H-h \right)}^{2}}}{\pi \cdot P_{r}^{\left( i \right)}}}.
\end{equation}
Given $d_{i}$, the horizontal distance is estimated as
\begin{align}\nonumber
{{r}_{i}}&=\sqrt{{{d}_{i}}^{2}-{{\left( H-h \right)}^{2}}}\\\label{eq9}
&=\sqrt{\sqrt{\frac{A{{T}_{s}}\left( \psi  \right)g\left( \psi  \right){{P}_{t}}{{\left( H-h \right)}^{2}}}{\pi {{P}_{r}}}}-{{\left( H-h \right)}^{2}}}.
\end{align}
The receiver coordinates (i.e., $\left( x,y \right)$), can be then obtained using following equations:

\begin{equation}
 \begin{cases}
  {{\left( x-{{x}_{1}} \right)}^{2}}+{{\left( y-{{y}_{1}} \right)}^{2}}=r_{1}^{2}& \\
 {{\left( x-{{x}_{2}} \right)}^{2}}+{{\left( y-{{y}_{2}} \right)}^{2}}=r_{2}^{2}& \\
 {{\left( x-{{x}_{3}} \right)}^{2}}+{{\left( y-{{y}_{3}} \right)}^{2}}=r_{3}^{2}& \\
 {{\left( x-{{x}_{4}} \right)}^{2}}+{{\left( y-{{y}_{4}} \right)}^{2}}=r_{4}^{2}& \\
\end{cases},
\end{equation}
where $({{x}_{i}},{{y}_{i}})$ are the horizontal coordinates of the $i^\text{th}$ transmitter decoded from the LED ID. Subtracting the last three equations from the first one, the following group of equations is obtained:
\begin{equation}\label{eq12}
\begin{cases}
\left( {{x}_{1}}-{{x}_{2}} \right)x+\left( {{y}_{1}}-{{y}_{2}} \right)y&\\
=\left( r_{2}^{2}-r_{1}^{2}-x_{2}^{2}+x_{1}^{2}-y_{2}^{2}+y_{1}^{2} \right)/2 &\\
 \left( {{x}_{1}}-{{x}_{3}} \right)x+\left( {{y}_{1}}-{{y}_{3}} \right)y&\\
 =\left( r_{3}^{2}-r_{1}^{2}-x_{3}^{2}+x_{1}^{2}-y_{3}^{2}+y_{1}^{2} \right)/2 &\\
 \left( {{x}_{1}}-{{x}_{4}} \right)x+\left( {{y}_{1}}-{{y}_{4}} \right)y&\\
 =\left( r_{4}^{2}-r_{1}^{2}-x_{4}^{2}+x_{1}^{2}-y_{4}^{2}+y_{1}^{2} \right)/2 &\\
\end{cases},
\end{equation}
Eq. (\ref{eq12}) can be written in matrix format as $\mathbf{AX}=\mathbf{B}$ where
\begin{equation}
\mathbf{A}=\left[ \begin{matrix}
   {{x}_{2}}-{{x}_{1}} & {{y}_{2}}-{{y}_{1}}  \\
   {{x}_{3}}-{{x}_{1}} & {{y}_{3}}-{{y}_{1}}  \\
   {{x}_{4}}-{{x}_{1}} & {{y}_{4}}-{{y}_{1}}  \\
\end{matrix} \right],
\end{equation}
\begin{equation}
\mathbf{B}=\frac{1}{2}\left[ \begin{matrix}
   \left( r_{1}^{2}-r_{2}^{2} \right)+\left( x_{2}^{2}+y_{2}^{2} \right)-\left( x_{1}^{2}+y_{1}^{2} \right)  \\
   \left( r_{1}^{2}-r_{3}^{2} \right)+\left( x_{3}^{2}+y_{3}^{2} \right)-\left( x_{1}^{2}+y_{1}^{2} \right)  \\
   \left( r_{1}^{2}-r_{4}^{2} \right)+\left( x_{4}^{2}+y_{4}^{2} \right)-\left( x_{1}^{2}+y_{1}^{2} \right)  \\
\end{matrix} \right],
\end{equation}
\begin{equation}
\mathbf{X}={{[x\ y]}^{T}}.
\end{equation}
Applying linear least square estimation approach \cite{14}, the final coordinates are estimated as
\begin{equation}
\mathbf{X}={{({{\mathbf{A}}^{T}}\mathbf{A})}^{-1}}{{\mathbf{A}}^{T}}\mathbf{B}
\end{equation}
\section{SIMULATION AND ANALYSIS}
In order to analyze multipath reflections in a typical room, three positions are selected. \emph{A} (0.01, 0.01) represents the corner point of the room where strong reflections and scatterings happen; \emph{B} (3, 0.01) represents the edge point where the reflections become medium; \emph{C} (3, 3) represents the point at the center of the room where the effect of reflections is weak.
\subsection{Impulse response analysis}
\begin{figure}
\centering
\includegraphics[width = 8cm, height = 7.5cm]{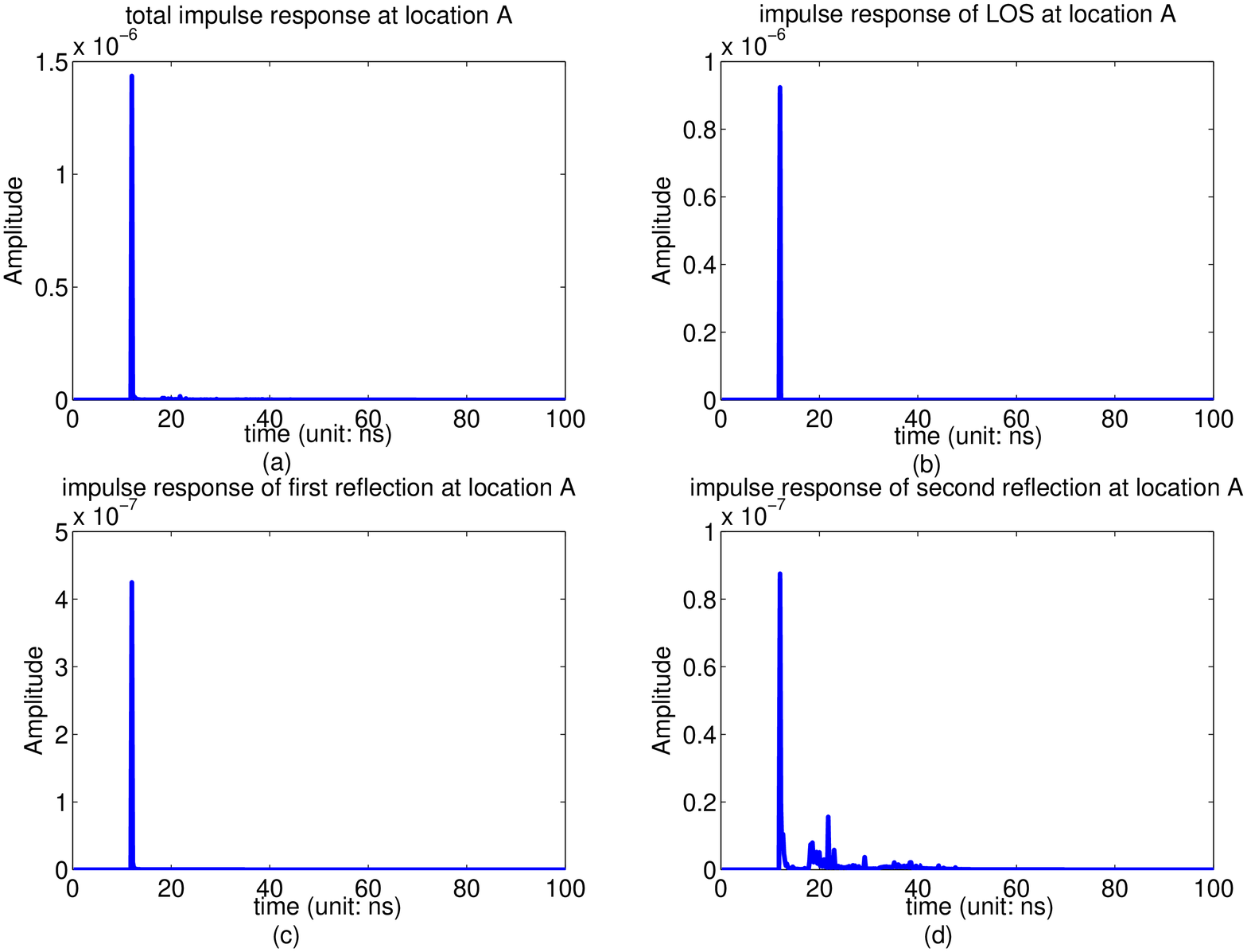}
\caption{Impulse response of each order of reflections at location \emph{A}.}
\end{figure}
Fig. 4 shows the impulse response at Location \emph{A}. The total impulse response is composed of the LOS, primary reflections and secondary reflections. Impulse response amplitude of the first reflections is comparable to that of the LOS inducing large positioning errors. Compared with the first reflections, the impulse response of secondary reflections has a smaller amplitude  while contributing more delay due to an elongated tail.
\begin{figure}
\centering
\includegraphics[width = 8cm, height = 7.5cm]{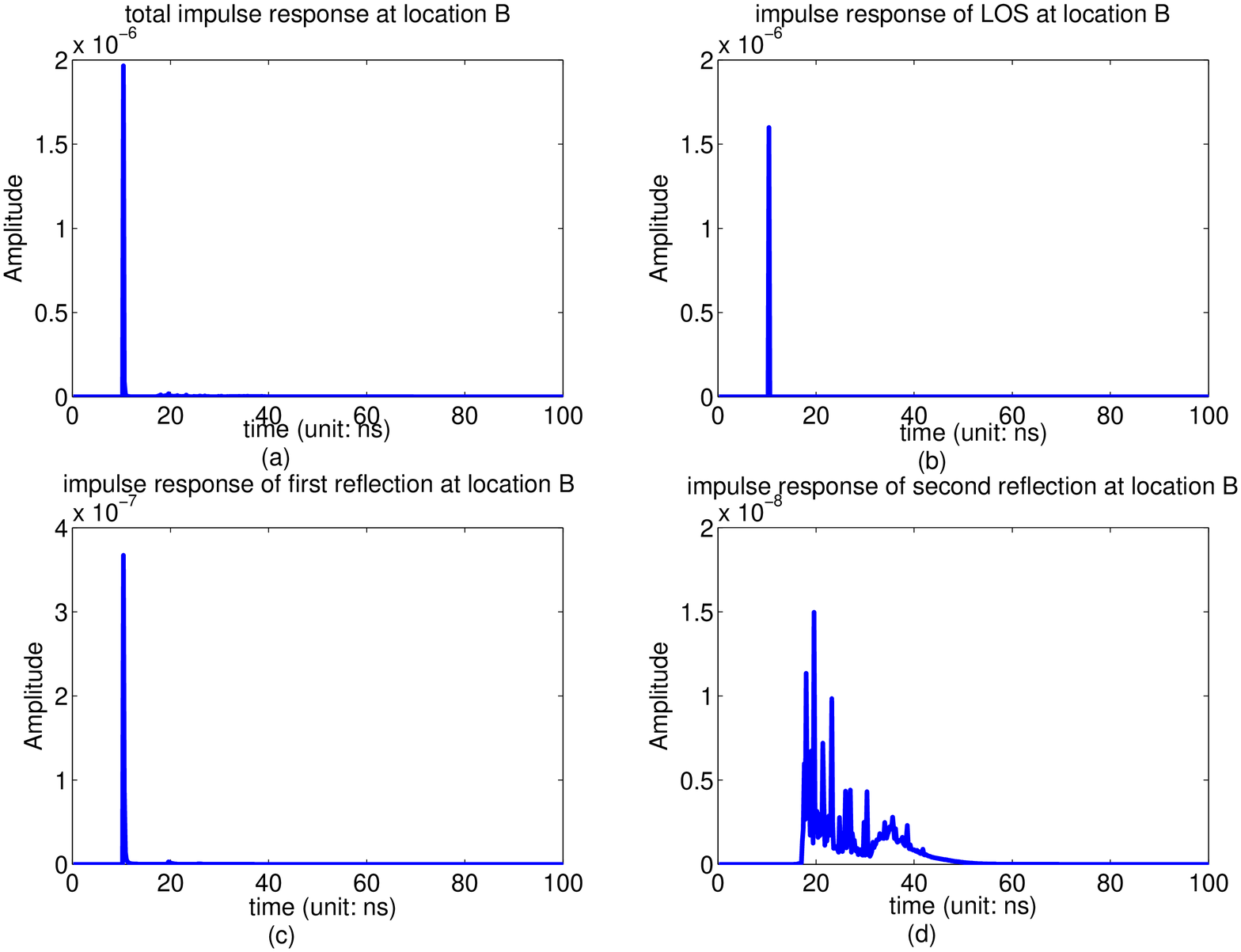}
\caption{Impulse response of each order of reflections at location \emph{B}.}
\end{figure}

Fig. 5 shows the impulse response at Location \emph{B}. Impulse response amplitude from the first reflections is much smaller than that of LOS. For the second reflections, the impulse response amplitude is much smaller than the first reflections. Thus, at this location, as we will show later, the effect of reflections on the positioning accuracy becomes less than Location \emph{A}.
\begin{figure}
\centering
\includegraphics[width = 8cm, height = 7.5cm]{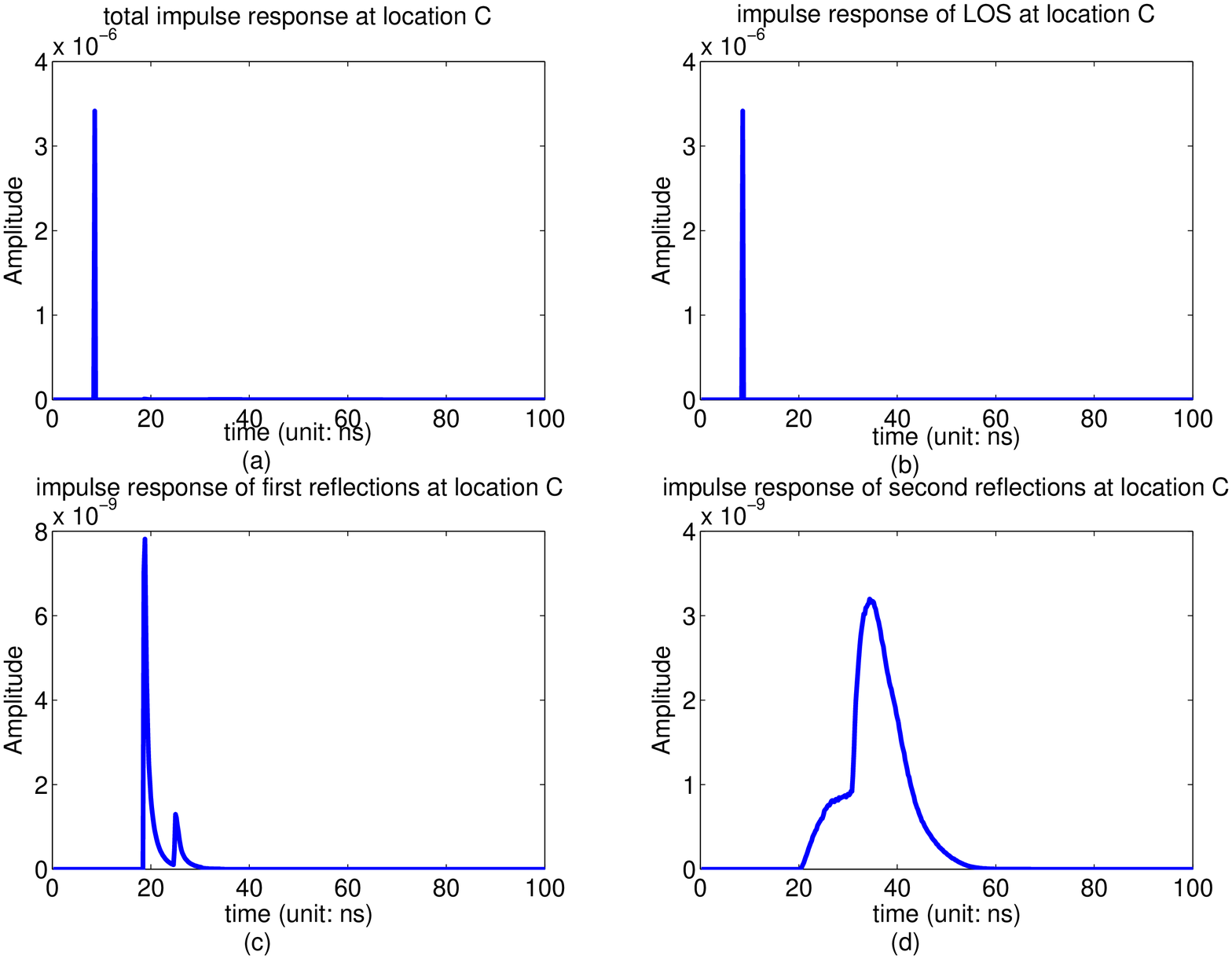}
\caption{Impulse response of each order of reflections at location \emph{C}.}
\end{figure}

Fig. 6 shows the impulse response at Location \emph{C} which is the center of the room. Compared to the LOS, the amplitudes from both reflections are small and can be ignored. Therefore, the effect of reflections on the positioning performance is expected to be weak in the central area.
\subsection{Power intensity distribution analysis}
As RSS information is applied in the positioning algorithm to estimate the distance between transmitter and receiver, the power intensity distribution affects the positioning accuracy directly. Time division multi-access (TDM) is assumed as channel access method; therefore, the four LED bulbs transmit signals at different time slots in one frame period \cite{15}. In this way, PD detects the power intensity from only one transmitter at every moment. Considering the symmetrical installation of four LED transmitters (TXs), power intensity distribution of the bulb located at (2, 2) is estimated where the transmitted power is 5 W. The received power of each order is shown in Fig. 7 through Fig. 9.

\begin{figure}
\centering
\includegraphics[width = 8cm, height = 7.5cm]{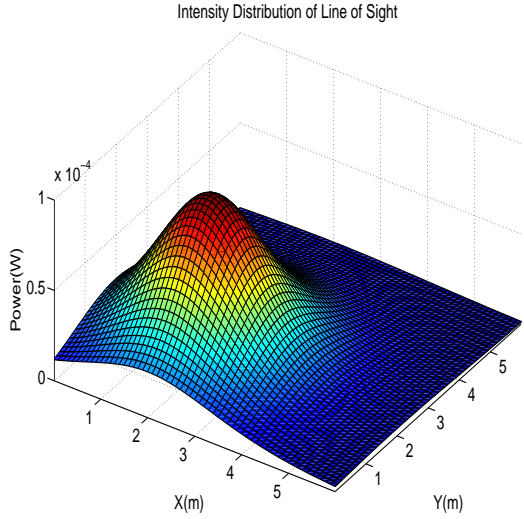}
\caption{Intensity distribution of LOS.}
\end{figure}

Particularly, Fig. 7 shows the power intensity distribution from LOS. The location which is right below the LED bulb has the highest received power. The received power at each location is inversely proportional to the distance-squared from the source  which is in correspondence with the power attenuation expressed as Eq. (\ref{eq6}).

\begin{figure}
\centering
\includegraphics[width = 8cm, height = 7.5cm]{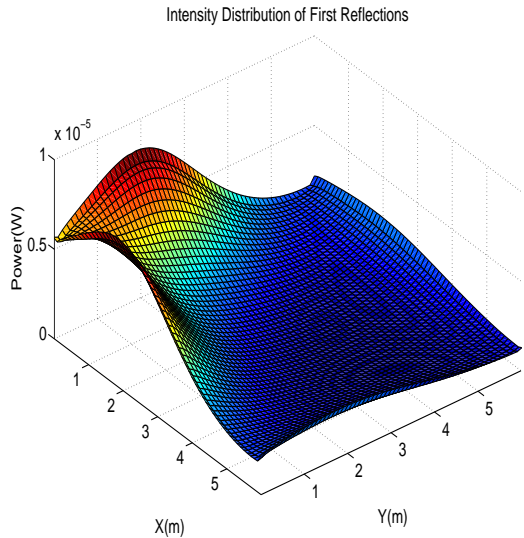}
\caption{Intensity distribution of first reflections.}
\end{figure}
Fig. 8 shows the power intensity distribution contributed by the first reflections. The power intensity is high at the edge and corner area near the transmitter's side where the reflections are strong. The reflected power will induce positioning errors since in the distance estimation, i.e., Eq. (\ref{eq6}), only direct power attenuation from the transmitter is considered.

\begin{figure}
\centering
\includegraphics[width = 8cm, height = 7.5cm]{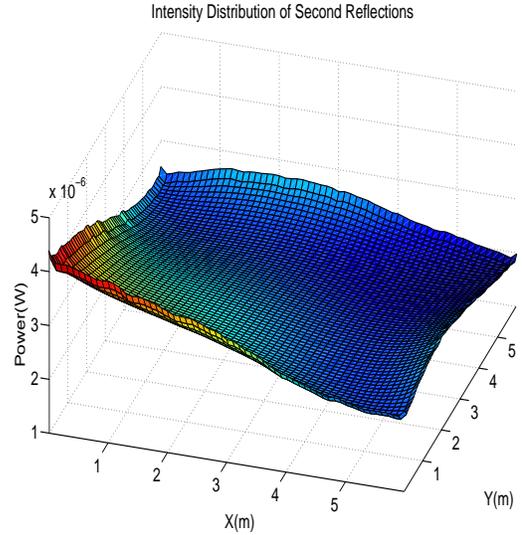}
\caption{Intensity distribution of second reflections.}
\end{figure}
Fig. 9 shows the power intensity distribution contributed by the second reflections. In the edge and corner area, the intensity is slightly higher while the total power intensity distribution becomes more uniform. Although the received power decreases a lot, these reflections still create positioning errors.

\begin{figure}
\centering
\includegraphics[width = 8cm, height = 7.5cm]{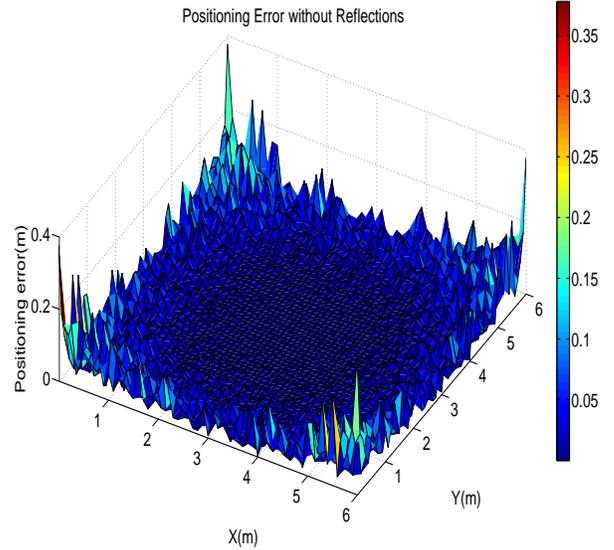}
\caption{ Positioning error without considering the reflections.}
\end{figure}
\subsection{Analysis of positioning accuracy}
With the RSS information and positioning algorithm, the receiver at each location is able to estimate its coordinates. As a benchmark and in order to show the effect of multipath reflections on the positioning accuracy, positioning error neglecting the reflected power is also calculated and shown in Fig. 10. As it can be seen, the positioning error is low all over the room, and only a little higher in the corner area because of the low signal-to-noise ratio (SNR).

\begin{figure}
\centering
\includegraphics[width = 8cm, height = 7.5cm]{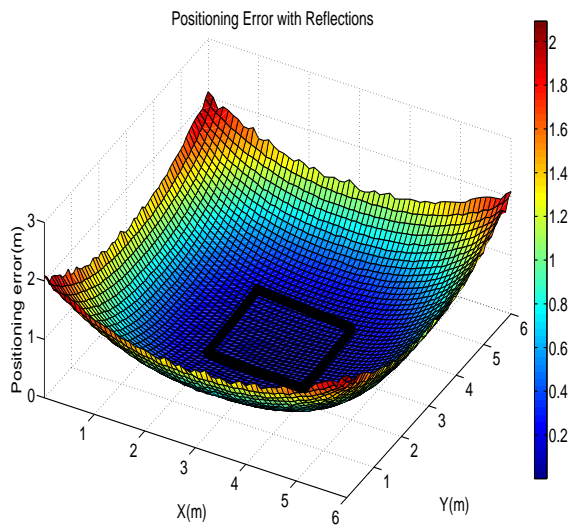}
\caption{Positioning error considering the reflections.}
\end{figure}
Fig. 11, on the other hand, shows the positioning performance considering the multipath reflections. The positioning performance is satisfactory within the rectangular shown in Fig. 11 where the LED bulbs are located right above the corners. For the central area, the positioning error is low, due to the small amplitude of impulse response as well as low power intensity contributed from reflections. The edge area has relatively higher error as the effect of the multipath reflections increases. The system has the worst performance at the corner area as the reflections are strongest there.

Fig. 12 and Fig. 13 present the histogram of the positioning error of the systems neglecting and considering the reflections, respectively. If no reflections are considered, the errors only come from the thermal noise and shot noise \cite{16} which are small. The reflections are major concerns in the system impairing significantly the system performance.

Finally, Table 2 compares the positioning error for the two cases of neglecting and considering the reflections. When reflections are neglected, the root mean square (RMS) error is 0.0423 m, and 0.1037  in the corner point. However, this situation is not practical as reflections always exist on the walls, ceiling and floor in a typical room.

\begin{figure}
\centering
\includegraphics[width = 8cm, height = 7cm]{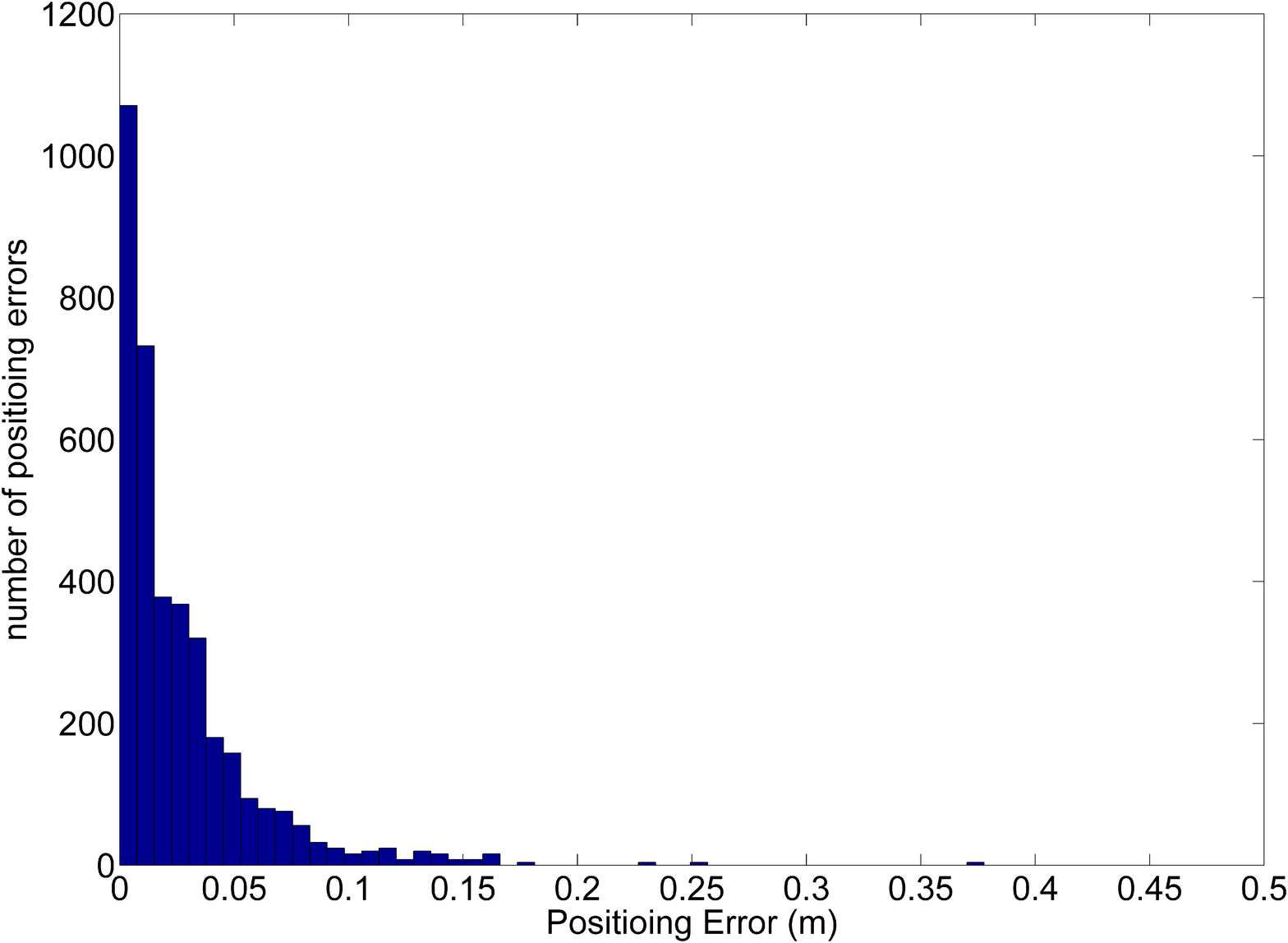}
\caption{ Histogram of positioning error with no reflections.}
\end{figure}
\begin{figure}
\centering
\includegraphics[width = 8cm, height = 7cm]{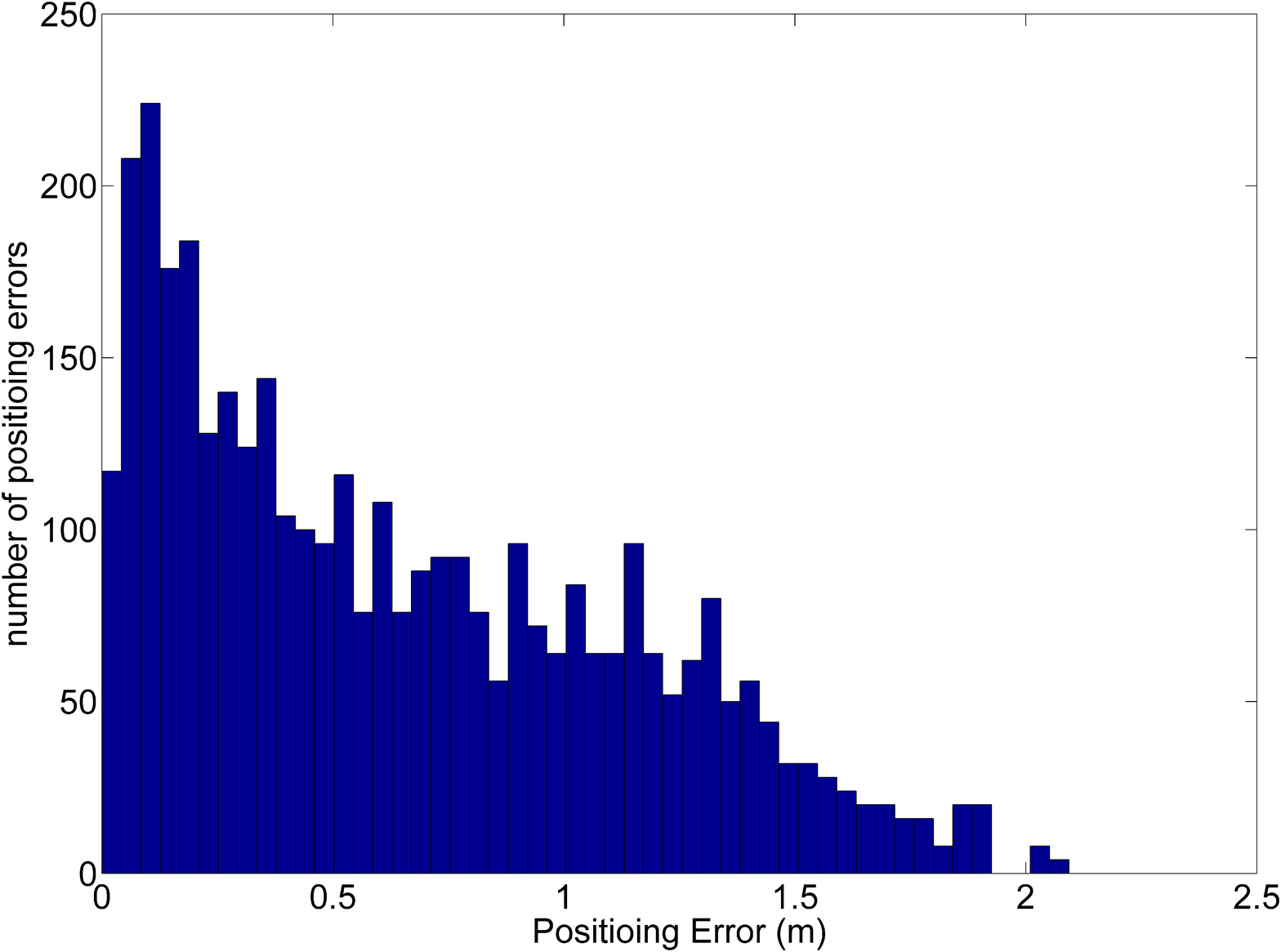}
\caption{Histogram of positioning error considering the reflections.}
\end{figure}
On the other hand, when multipath reflections are taken into account, the positioning accuracy decreases differently in  distinct locations. At Location \emph{A} and \emph{B}, the positioning errors are dramatically increased to 2.0989 m and 1.2772 m, respectively. As expected, the positioning error at location \emph{B} is less than location \emph{A} since the reflections at the edge are not as strong as those at the corner. However, at the central point (i.e., Location \emph{C}), the positioning performance is almost not affected by the reflections. For the rectangular area covered perfectly by the four LED bulbs, the RMS error is 0.0676 m which is satisfactory for indoor custom application. The total RMS error is 0.8064 m, as the rectangular area covered by the LED bulbs is only 11.1\% of the total area. Layout design of the LED bulbs should be addressed in the future work to increase the average positioning accuracy.
\begin{table}
\begin{center}
\begin{quote}
\caption{Positioning error with/without reflections}
\label{table2}
\end{quote}
\begin{tabular}{|c|c|c|}
  \hline
   &\textbf{Reflections}   &\textbf{Reflections}\\
  &\textbf{neglected (m)} & \textbf{included (m)}\\ \hline
  Loc. \emph{A}  &0.1037 &2.0989 \\ \hline
  Loc. \emph{B}  &0.0075 &1.2772 \\ \hline
  Loc. \emph{C}  &0.0017 &0.0019 \\ \hline
  RMS error of  &0.0031 &0.0676 \\
  the rectangular area & &\\\hline
  RMS error of   &0.0423 &0.8064 \\
  the whole room & & \\\hline
  \end{tabular}
\end{center}
\end{table}

\section{CONCLUSIONS}
In this paper, an indoor visible light positioning system taking into account multipath reflections has been investigated for a typical room where the impulse response is obtained employing CDMMC approach. In particular, the impulse response of each order has been calculated for three locations representing the corner, edge and center points. As RSS information is applied to calculate the distance between a transmitter and receiver, the power intensity distribution contributed by each reflection has also been computed and shown. Finally, positioning errors in all the locations of the room have been estimated and compared with those of previous studies  where reflection is neglected. We have shown that the corner area suffers severely from the reflections, while  the impact of reflections decrease in the edge areas. It has been also shown that the central area is almost unaffected by the multipath reflections. For future work, calibration methods should be considered to increase the positioning accuracy in the corner and edge areas, taking into account the multipath reflections. In addition, by increasing the coverage area of the LEDs with a suitable layout design, the average positioning error can be decreased.
\section*{Acknowledgement}
The authors would like to thank the National Science Foundation (NSF) ECCS directorate for their support of this work under Award \# 1201636, as well as Award \# 1160924, on the NSF “Center on Optical Wireless Applications (COWA– \href{http://cowa.psu.edu}{http://cowa.psu.edu})”
\balance
\bibliographystyle{IEEEtran}
\bibliography{ref}

\end{document}